\documentstyle[prl,multicol,epsf,epsfig,aps]{revtex}
\newcommand{\BEQ}{\begin{equation}}
\newcommand{\EEQ}{\end{equation}}
\newcommand{\BEA}{\begin{eqnarray}}
\newcommand{\EEA}{\end{eqnarray}}

\renewcommand{\d}{{\rm d}}

\newcommand{\ri}{{\rm i}} 
\renewcommand{\thesection}{\arabic{section}}

\begin{document}
\draft
\twocolumn
\title{Arrows of Time and Chaotic Properties of the Cosmic Background
Radiation}

\author{A.E. Allahverdyan$^{1,2)}$, V.G. Gurzadyan$^{2,3)}$}
\address{ $^{1)}$ 
Service de Physique Th\'eorique, CEA/DSM/SPhT,\\
Unit\'e associ\'ee au CNRS, CEA/Saclay,
91191 Gif-sur-Yvette cedex, France\\
$^{2)}$Yerevan Physics Institute, Alikhanian Brothers St. 2, Yerevan 
and Garni Space Astronomy Institute, Garni,
Armenia,\\
$^{3)}$ ICRA, Dipartimento di Fisica, Universita di
Roma La Sapienza, Rome, Italy
}

\maketitle

\begin{abstract}
  
  We advance a new viewpoint on the connection between the the
  thermodynamical and cosmological arrows of time, which can be traced
  via the properties of Cosmic Microwave Background (CMB) radiation.
  We show that in the Friedmann-Robertson-Walker Universe with
  negative curvature there is a necessary ingredient for the existence
  of the thermodynamical arrow of time.  It is based on the dynamical
  instability of motion along null geodesics in a hyperbolic space.
  Together with special (de-correlated) initial conditions, this
  mechanism is sufficient for the thermodynamical arrow, whereas the
  special initial conditions alone are able to generate only a
  pre-arrow of time.  Since the negatively curved space will expand
  forever, this provides a direct connection between the
  thermodynamical and cosmological arrows of time. The structural
  stability of the geodesic flows on hyperbolic spaces and hence the
  robustness of the proposed mechanism is especially stressed.  We
  then point out that the main relations of equilibrium statistical
  thermodynamics (including the second law) do not necessarily depend
  on any arrow of time.  Finally we formulate a {\it curvature
  anthropic principle}, which stipulates the negative curvature as a
  necessary condition for the time asymmetric Universe with an
  observer. CMB has to carry the signature of this principle as well.

\end{abstract}


\renewcommand{\thesection}{\arabic{section}}
\section{ Introduction}
\setcounter{equation}{0}\setcounter{figure}{0} 
\renewcommand{\thesection}{\arabic{section}.}

The structure and mechanisms of time-asymmetry, or the arrows of time,
remain among the much debated questions of the modern physics, in
spite of the intense attention devoted to these topics over many
years.  The relation between the thermodynamical and the cosmological
arrow has been studied from very different viewpoints, e.g. by Gold,
Penrose, Hawking \cite{Gold}|\cite{Hawking}, Page \cite{Page},
Petrosky
and Prigogine \cite{PP}, Zeh and others \cite{perez,Zeh}.

In the present paper we argue that the cosmological arrow of time, as
defined by the expansion of the universe, and a crucial observational
fact on the existence of highly isotropic Cosmic Microwave Background
radiation (CMB) with the Planckian spectrum, can be connected with the
thermodynamical arrow of time; preliminary account of these views are
in
\cite{AG}  We will argue that the curvature of
the universe can have an essential role in this problem. CMB is a
cornerstone of our discussion \cite{solvay}.  CMB photons are moving
freely during almost the entire lifetime of the Universe, thus tracing
its geometry.  Indeed, the properties of the CMB in a hyperbolic
Friedmann-Robertson-Walker $(k=-1)$ Universe differ from those of
flat, $k=0$, and positively curved, $k=+1$, cases.

In particular, the exponential deviation of the geodesics and the
ensuing
effect of geodesic mixing in the hyperbolic spaces leads at least to
the
following observable consequences \cite{g1,g2}:
\\
(1) damping of anisotropy after the last scattering epoch;
\\
(2) flattening of the autocorrelation function;
\\
(3) distortion of anisotropy spots.

Statistically significant signature of the third effect | the
threshold independent ellipticity in the CMB sky maps | has been
detected for COBE-DMR 4-year data \cite{g3}.  Interpreted as a result
of the geodesic mixing of photons, as predicted \cite{g2}, this will
model-independently indicate the negative curvature of the Universe
and $\Omega_0 < 1$. More advanced descriptor to trace the curvature is
the Kolmogorov complexity of CMB anisotropies \cite{Gu}, \cite{AGS}.

The recent Boomerang \cite{Bern}, Maxima, DASI, CPI (see \cite{CPI})
data on the CMB power spectrum are interpreted as supporting the CDM
flat models with adiabatic scale-invariant fluctuation spectrum,
though there are claims for other models as well, see e.g.
\cite{baker,WSP,mc}.  The precise flatness however cannot be proved
not only due to the measurement errors but mainly due to the
degeneracy and dependence on a number of free parameters.

It is remarkable that threshold independent behaviour of the
ellipticity of the
anisotropies has been found also for Boomerang maps \cite{boom} which
differ from
COBE-DMR maps not only by their higher angular resolution but also
with lower noise level. 
The supernovae data, supporting the existence of a cosmological term,
are 
not decisive for the sign of the curvature \cite{perl,reiss}.

Let us now turn to the time asymmetry of the universe.  Several
different and at first glance, independent time-arrows have been
defined \cite{Penrose1}.\\
{\it (i)} Thermodynamical: entropy of a 
closed system increases with time. \\
{\it (ii)} Cosmological: the universe expands.\\ 
{\it (iii)} Psychological: knowledge of the past
but not of the future.\\
{\it (iv)} Electromagnetic: retarded interaction (propagation).\\
{\it (v)} Quantum-mechanical: the change of a wave function during 
the typical measurement process is irreversible i.e. neither unitary 
nor linear.

The electromagnetic and psychological arrows are viewed as a
consequence of the thermodynamical arrow
\cite{Penrose1,Penrose2,Zeh,davies}.  Following, in particular, Landau
and Lifshitz \cite{landau}, some researchers may think that the
quantum-mechanical arrow is independent of the others, and may even
serve as a base for them. In contrast, we believe that there is no
fundamental quantum-mechanical arrow of time, and the problem of
quantum measurement can be fully explained within quantum statistical
mechanics, i.e. essentially as a consequence of the thermodynamical
arrow of time \cite{a1}.  Notice also that we did not involve here the
CP-asymmetry of weak interactions, since it is not directly relevant
to the present discussion \cite{Penrose1}.

Finally, only the thermodynamical and cosmological arrows are basic
for our present purposes.  Broadly speaking, the thermodynamical arrow
for a statistical system can be formulated as a consequence of the
following {\it necessary} conditions:

1) De-correlated (special) initial conditions;

2) No-memory dynamics.

Depending on their background people are sometimes inclined to
overestimate
one of those reasons, thereby underestimating the other. However, it
should
be emphasized once more that {\it both of them} are strictly
necessary, 
as we show below.
The above two conditions appear already in the Boltzmann's derivation
of his
kinetic equation, though perhaps not explicitly.
They can be traced out clearly in Zwanzig's derivation of
master-equation \cite{zwanzig} or Jaynes' information-theoretical
approach to irreversibility \cite{jaynes}. A conventional discussion
about
possible relations between cosmological and thermodynamical arrows of
time
concentrates only the first condition \cite{Penrose1}|\cite{Page}, 
\cite{Zeh,davies}.

One of our main intentions in the present paper
is to show that this is not sufficient, because special initial
conditions 
alone can generate only a thermodynamical pre-arrow of time.
Our main purpose here is to point out that along with the initial
conditions, 
the 
second ingredient of the thermodynamical arrow can have a cosmological
context as well, which arises due to mixing of trajectories in
hyperbolic spaces.

Namely, if the Friedmann-Robertson-Walker universe has negative
curvature,
then the flow of null geodesics which describes the free motion of
photons, represents an Anosov system \cite{Anosov}, a class of
dynamical systems
with maximally strong statistical properties. Anosov systems are
characterized
by exponential divergence of initially close trajectories, by the
property of
K-mixing, positive Kolmogorov-Sinai (KS)-entropy, and countable
Lebesgue spectrum.
In particular, geodesic flows on compact manifolds with negative
constant 
curvature are characterized by the exponential decay of the
correlators 
\cite{Po}.
One of the significant properties of Anosov systems
is their structural stability, namely resistance of properties with
respect to perturbations. This is crucial, since we live not in a
universe
with strongly constant curvature but with small perturbations of
metric, and 
moreover, we know  the magnitude of their smallness from the same CMB
measurements.  

On the other hand, sufficiently fast decoupling of correlations is
responsible for the so-called Markovian behavior
\cite{arnold,zaslav,allah}.  As we shall discuss below this latter
property can be one of two main ingredients ensuring the
thermodynamical arrow, though the existence of other mechanisms is not
excluded.

Thus we will show that the exponential instability of geodesic flow in
hyperbolic FRW Universe | the geodesic mixing | revealed through the
properties of CMB, is relating the thermodynamical and cosmological
arrows of time.  The proposed mechanism does not offer an answer to an
ambagious question, what will happen with the thermodynamical arrow if
the cosmological arrow were to inverted, just because in the
negatively curved space the cosmological arrow will never be inverted.
One notices that many considerations and speculations on this question
implicitly identify the thermodynamical arrow of time with the second
law of thermodynamics, and appearance of a Gibbs distribution. In this
context we will show that the second law, and the Gibbs distribution
can be obtained from purely time-symmetric arguments, and need not be
consequences of the thermodynamical arrow. Further clarification is
needed to understand the link of cosmology with concrete aspects of
statistical physics.

The paper is organized as follows. In section \ref{thermo} we discuss
the thermodynamical arrow and the conditions which are necessary for
its derivation. In section \ref{geo} we show how the CMB mixing
properties in the negatively curved space can be connected with the
thermodynamical arrow.  Then we discuss the derivation of the second
law and the Gibbs distribution in a way that does not depend on the
thermodynamical arrow of time.  Our conclusions are presented in the
last section.

\renewcommand{\thesection}{\arabic{section}}
\section{Thermodynamical arrow of time}
\setcounter{equation}{0}\setcounter{figure}{0} 
\renewcommand{\thesection}{\arabic{section}.}
\label{thermo}
\subsection{Pre-arrow of time}

In the present section we will discuss the following aspects of the 
thermodynamical arrow of time. Starting from the standard system-bath
approach we will indicate how the choice of only special initial
conditions
leads to a thermodynamical {\it pre-arrow} of time. This is not 
sufficient to generate the full thermodynamical arrow. Under suitable 
dynamical conditions 
related to the features of the bath, the pre-arrow generates the full
thermodynamical arrow, namely a monotonous change with time of the
proper 
thermodynamical potential, which is most typically entropy or free
energy.
In general, the arrow is present only for a relatively later stage of
the 
relaxational dynamics of the system. Most frequently, this stage is
connected
with the markovian properties. However, in certain special situations
the
thermodynamical arrow may be present also in a non-markovian
situation.
Such a presentation of the thermodynamical arrow will, in particular,
make clear that in general {\it both} dynamical features and special 
initial conditions are necessary for establishing the thermodynamical 
arrow of time.

For the sake of generality and simplicity we will work within the 
quantum-mechanical formalism. The reader can just keep in mind that 
all our results are directly transportable to the classical physics 
upon changing density matrices to probability distributions, traces
to integration over the phase-space, and the
von Neumann equation to the Liouville equation.

Let a quantum system ${\rm S}$ interacts with a thermal bath ${\rm
B}$.
The total Hamiltonian is
\BEA
\label{mis1}
H=H_{\rm S}+H_{\rm B}+H_{\rm I}.
\EEA
We denote by $H_{\rm S}$ and $H_{\rm B}$
the Hamiltonians of the system and the bath respectively, whereas 
$H_{\rm I}$ stands for the interaction Hamiltonian.
The state of the full system is described by the density matrix ${\cal
D}(t)$,
which satisfies the corresponding von Neumann equation
\BEA
\label{mis2}
i\partial _t{\cal D}(t) =[H,{\cal D}(t)],\quad
{\cal D}(t)=e^{-itH/\hbar}{\cal D}(0)e^{itH/\hbar},
\EEA
where $[...,...]$ stands for the commutator as usual. 
The crucial assumption on the initial state can be formulated as
follows:
\BEA
\label{mis3}
{\cal D}(0)=D_{\rm S}(0)\otimes D_{\rm B}(0),
\EEA
which means that at the initial time $t=0$ the system and the bath
were completely independent. 
The state of the system at arbitrary positive time $t$ is described
by the corresponding partial density matrix:
\BEA
\label{rashid}
D_{\rm S}(t)={\rm tr}_{\rm B}{\cal D}(t),
\EEA
where ${\rm tr}_{\rm B}$ indicates the trace over the Hilbert space of
the 
bath. The important point of the system-bath approach | as well as any

statistical physics approach which derives the thermodynamical arrow |
is its dependence on incomplete observability: although the system and
the bath constitute a closed system, one is interested in the state of
the
system only, which under the presence of the bath evolves according to
a 
non-unitary dynamics generated by a superoperator ${\cal T}$:
\BEA
\label{khalil}
D_{\rm S}(t)={\cal T}(t,0)\,D(0)=
\sum_{\alpha\beta}A_{\alpha\beta}D(0)A^\dagger_{\alpha\beta},
\EEA
where $A_{\alpha\beta}$ are operators in the Hilbert space of ${\rm
S}$. 
They are determined via the spectral decomposition of the
initial density matrix of the bath
\BEA
\label{said}
D_{\rm B}(0)=\sum_\alpha\lambda_\alpha|\alpha\rangle
\langle \alpha|,\quad \langle \alpha|\beta\rangle
=\delta_{\alpha\beta},
\EEA
where Kronecker 
$\delta_{\alpha\beta}=1\,(0)$ for $\alpha=\beta$ ($\alpha\not
=\beta$),
and by the evolution operator generated by the complete Hamiltonian
$H$:
\BEA
\label{abdulla}
A_{\alpha\beta}=\sqrt{\lambda_{\beta}}\,
\langle \alpha|e^{-itH/\hbar}|\beta\rangle
\EEA
Eqs.~(\ref{khalil}, \ref{abdulla}) are easily obtained from
(\ref{mis2},
\ref{rashid}) upon substituting there (\ref{said}). One can check
directly 
that
\BEA
\sum_{\alpha\beta}A^\dagger_{\alpha\beta}A_{\alpha\beta}=1,
\EEA
as required for the trace-conservation of $D_{\rm S}(t)$ at any time.
Three important facts should be noticed in the context of
(\ref{khalil}): 
({\it i}) The superoperator ${\cal T}$ appearing in (\ref{khalil}) is
not
unitary, and in general it does not have an inverse operator. 
Thus, the dynamics of the
system alone is irreversible. This is a consequence of the general
fact that
statistical systems are described incompletely, e.g. in the
system-bath 
approach one focusses on the system alone in the presence of the bath.

({\it ii}) The operators $A_{\alpha\beta}$ do not depend on the
initial state
of the system itself. Thus, the dynamics of the system is {\it
autonomous},
solely due to the initial condition (\ref{mis3}). It is obvious that
this
property will not be valid for an arbitrary initial state.
({\it iii}) In general, the property ${\cal T}(t_{f},t_i)
={\cal T}(t_{f}-t_i)$ for all $t_f> t_i$, which is automatically valid
for 
the unitary situation, is broken inasmuch as the bath is present.

It appears that Eqs.~(\ref{mis2}, \ref{mis3}) are enough to ensure the
existence of the pre-arrow, which is not a statement on the dynamics
of the 
system itself but rather a statement on the similarity between the
dynamical processes given by (\ref{khalil}) and that generated by the
same 
Hamiltonian (\ref{mis1}) and somewhat different initial condition:  
\BEA
\label{mis4}
{\cal R}(0)=R_{\rm S}(0)\otimes D_{\rm B}(0).
\EEA
Notice that the difference between (\ref{mis3}) and (\ref{mis4}) is
only
in the initial condition for the system itself: $R_{\rm S}(0)
\not =D_{\rm S}(0)$. 
An important measure of difference between 
$R_{\rm S}(0)$ and $D_{\rm S}(0)$ is the relative entropy
\cite{balian}:
\BEA
\label{rel}
S[D_{\rm S}(0)\,||\,
R_{\rm S}(0)]={\rm tr}[
D_{\rm S}(0)\ln D_{\rm S}(0)-D_{\rm S}(0)\ln R_{\rm S}(0)],\nonumber
\EEA
which is known to be non-negative and is equal to zero only for
$R_{\rm S}(0)
=D_{\rm S}(0)$. In general, the relative entropy 
$S[D_{\rm S}(0)\, ||\,R_{\rm S}(0)]$
characterizes the information needed to distinguish between 
the density matrices $D_{\rm S}(0)$ and $R_{\rm S}(0)$
via many ($\gg 1$) independent identical experiments \cite{sh}.
The fundamental theorem \cite{lindblad,sh,allah} states that at all
later
times the relative entropy between the density matrices 
$R_{\rm S}(t)={\rm tr}_{\rm B}{\cal R}(t)$
and $D_{\rm S}(t)={\rm tr}_{\rm B}{\cal D}(t)$
does not increase:
\BEA
S[D_{\rm S}(0)\,||\,
R_{\rm S}(0)]&&\ge S[D_{\rm S}(t)\,||\,
R_{\rm S}(t)]
\nonumber\\
\equiv&& S[{\cal T}(t,0)\,D_{\rm S}(0)\,||\,
{\cal T}(t,0)\,R_{\rm S}(0)].
\label{lindblad}
\EEA
The equality sign in (\ref{lindblad}) is realized for unitary 
evolution showing that there is no pre-arrow for a closed system.
Equation (\ref{lindblad}) shows that any dynamics for the system 
with the initial conditions (\ref{mis3}) does not increase the
distinguishability between different initial conditions.

\subsection{Arrow of time}

Two additional dynamical conditions which lead to the appearance of
the 
thermodynamical arrow of time are the following: 1) Features of the
bath are 
such that for sufficiently large $t$ one has
\BEA
\label{baba}
{\cal T}(t,0)={\cal T}(t),
\EEA
i.e. the dependence on the initial time disappears (no-memory).
2) The system relaxes with time to a certain stationary density matrix

$D^{(st)}_{\rm S}$:
\BEA
\label{handi}
{\cal T}(t)D^{(st)}_{\rm S}=D^{(st)}_{\rm S},\quad
D_{\rm S}(t)\to D^{(st)}_{\rm S}.
\EEA
For times where (\ref{baba}) is valid, one can apply (\ref{lindblad}) 
for any $\theta$ as
\BEA
S[D_{\rm S}(t)\,||\,
D^{(st)}_{\rm S}]&&\ge 
S[{\cal T}(\theta) D_{\rm S}(t)\,||\,{\cal T}(\theta)
D^{(st)}_{\rm S}]
\nonumber\\
&&\equiv S[D_{\rm S}(t+\theta)\,||\,
D^{(st)}_{\rm S}],
\label{saladin}
\EEA
and deduce that the function $S[D_{\rm S}(t)\,||\,
D^{(st)}_{\rm S}]$
is monotonically decreasing with time, since $\theta>0$ was arbitrary.

The concrete properties of this
function depends on the structure of $D^{(st)}_{\rm S}$. If, for
example,
the stationary distribution is microcanonical:
$D^{(st)}_{\rm S}\propto 1$, then (\ref{saladin}) reduces to the
statement that
von Neumann entropy $$S_{vN}[D_{\rm S}(t)]=-{\rm tr}[D_{\rm S}(t)\ln 
D_{\rm S}(t)]$$ increases with time. In the case of the canonical
distribution
function $D^{(st)}_{\rm S}\propto \exp(-H_{\rm S}/T)$, where $T$ is
temperature, one gets that the free energy $$F=U(t)-TS_{vN}(t),
\quad U(t)={\rm tr}[D_{\rm S}(t)H_{\rm S}]$$ monotonically decreases
with
time. Here $U(t)$ is the average energy; if it is conserved during
evolution, 
then the statements on free energy and entropy are essentially
equivalent.
Notice the difference with the pre-arrow of time which compares only 
the initial relative entropy $S[D_{\rm S}(0)\,||\,
R_{\rm S}(0)]$
with the relative entropy $S[D_{\rm S}(t)\,||\,
R_{\rm S}(t)]$ at any
time $t>0$, without making any connection between 
$S[D_{\rm S}(t)\,||\,
R_{\rm S}(t)]$ and $S[D_{\rm S}(t+\theta)
\,||\,
R_{\rm S}(t+\theta)]$ for $\theta>0$.

To summarize this subsection, we notice that the conditions 
(\ref{baba}) ensuring the appearance of the thermodynamical
arrow of time is only sufficient; in certain situations it can be
substituted by other (weaker) dynamical assumptions. 
The relaxation (\ref{handi}) should be, of course, understood on times
much less than the Poincar\'e recurrent time.
Therefore, the recurrent time itself must be very large. This
condition can
be considered as satisfied, since for majority of ``reasonable''
systems
the Poincar\'e time exceeds the age of the Universe.

\subsection{A scenario for the no-memory regime}

In the present section we will discuss a possible scenario for the
appearance of the 
condition (\ref{baba}). It is based on a sufficiently weak coupling
between the system and the bath, as well as on the fast relaxation of
the
bath correlation functions. This last feature is intrinsic for the
bath, and (for the considered limit) it has nothing to do with the
coupling
to the system. Starting from (\ref{mis1}), it is convenient to
introduce Liouville (super)operators:
\BEA
&&{\cal L}_{\rm k}(t)=\frac{1}{\ri\hbar}[H_{\rm k},...],\quad
{\rm k=S,B},\\
&&{\cal L}_{\rm I}(t)=\frac{1}{\ri\hbar}[H_{\rm I}(t),...]
\EEA
where $H_{\rm I}(t)$ is the corresponding Heisenberg operator in the
free
representation (i.e. without the interaction):
\BEA
H_{\rm I}(t)=
e^{\frac{\ri t}{\hbar}\,(H_{\rm S}+H_{\rm B}\,)}
\,H_{\rm I}\,e^{-\frac{\ri t}{\hbar}\,(H_{\rm S}+H_{\rm B}\,)}.
\EEA
Using Liouville operators, the full dynamics for the overall density
matrix ${\cal D} (t)$ is written as
\begin{eqnarray}
{\cal D} (t) =e^{t\,(
{\cal L}_{\rm S}+{\cal L}_{\rm B})}\,
{\rm \hat{T}}\,e^{\int_0^t\d \theta
{\cal L}_{\rm I}(\theta)}\, D_{\rm B}(0)\otimes D_{\rm S}(0),
\end{eqnarray}
where ${\rm \hat{T}}$ is the time-ordering operator, and where
$D_{\rm B}(0)$ and $D_{\rm S}(0)$ refer to the initial density
matrices of 
the  bath and the system respectively. The marginal density matrix 
of the system, $D_{\rm S}(t)={\rm tr}_{\rm B}{\cal D}(t)$, reads:
\BEA
D_{\rm S}(t)=e^{t{\cal L}_{\rm I}}\,\langle 
{\rm \hat{T}}\,e^{\int_0^t\d \theta\,
{\cal L}_{\rm I}(\theta)}\rangle\,D(0),
\label{basu}
\EEA
where for any quantity ${\cal X}$ (possibly a superoperator):
\BEA
\langle {\cal X}\rangle\equiv{\rm tr}_{\rm B}[ {\cal X}D _{\rm B}(0)].
\EEA
Notice that the mutual ordering between ${\cal X}$ and $D_{\rm B}(0)$
can be important.
By analogy with the classical cumulant expansion one writes
\cite{kubo}
\BEA
\label{ss}
\langle{\rm \hat{T}}\,e^{\int_0^t\d \theta\,
{\cal L}_{\rm I}(\theta)}\rangle=
{\rm \hat{T}}\,e^{\int_0^t\d \theta\,{\cal F}(\theta)},
\EEA
where ${\cal F}$ is another superoperator, which is determined
step-by-step
by expanding both sides of (\ref{ss}) over $H_{\rm I}$:
\BEA
{\cal F}=\sum_{k=1}{\cal F}_k,
\nonumber
\EEA
with
\BEA
&&\langle{\rm \hat{T}}\,e^{-\frac{\ri}{\hbar}\int_0^t\d \theta\,
{\cal L}_{\rm I}(\theta)}\rangle=
1+\\
&&\sum_{k=1}\int_0^t\d \theta_1\int_0^{\theta_1}
\d \theta_2......\int_0^{\theta_{k-1}}
\d \theta_k\, \langle
{\cal L}_{\rm I}(\theta_1)...{\cal L}_{\rm I}(\theta_k)
\rangle.\nonumber
\EEA
The first two contributions into ${\cal F}$ are the following:
\BEA
&&{\cal F}_1(t)=\langle {\cal L}_{\rm I}(t)\rangle,\\
&&{\cal F}_2(t)=\int_0^t\d \theta\left[
\langle {\cal L}_{\rm I}(t){\cal L}_{\rm I}(\theta)\rangle -
\langle {\cal L}_{\rm I}(t)\rangle\langle{\cal L}_{\rm I}(\theta)
\rangle\right]\nonumber.
\EEA
The content of the considered approximation is that one keeps only
these two terms for ${\cal F}$, thus neglecting all other cumulants.
This is applicable if the magnitude of $H_{\rm I}$ is sufficiently
small.
The cumulant expansion also ensures that possible secular terms are
absent,
so that the neglected higher-order cumulants are typically
homogeneously small 
compared with the second one \cite{kubo}.
Let us assume for simplicity that 
\BEA
\label{abbat}
{\cal F}_1(t)=0, 
\EEA
and then the final differential convolutionless equation 
for $D_{\rm S}(t)$ reads from (\ref{basu}):
\BEA
\label{bfinal}
&&\dot{D}_{\rm S}(t)=\frac{1}{\ri\hbar}\,[H_{\rm S}(t),D_{\rm S}(t)]+
e^{t{\cal L}_{\rm S}}\,{\cal F}_2(t)\,
e^{-t{\cal L}_{\rm S}}D_{\rm S}(t).\nonumber\\
&&
\EEA
Notice that this is a differential, though non-markovian, equation for
$D(t)$. So it is consonant with thermodynamical pre-arrow of time, but
may be compatible with a non-monotonic change of the corresponding
thermodynamical potential.  To implement the no-memory approximation,
we will work out a particular case, where the initial interaction
Hamiltonian is presented as: \BEA H_{\rm I}=S\otimes B, \EEA with $S$
and $B$ belonging to the Hilbert spaces of the system and the bath
respectively. Then Eq.~(\ref{bfinal}) reads: \BEA
\label{bresult}
&&\dot{D}_{\rm S}(t)=\frac{1}{i\hbar}[H_{\rm S},D_{\rm S}(t)]- 
\frac{1}{\hbar^2}\int_0^t\d \theta\, \{\,K(t,\theta)\times\\
&&(\,S\,S(\theta-t)D_{\rm S}(t)-
\,S(\theta-t)D_{\rm S}(t)S\,)\, +{\rm h.c.}\,\},\nonumber\\
&&S(t)=e^{itH_{\rm S}/\hbar}Se^{-itH_{\rm S}/\hbar},\\
&&B(t)=e^{itH_{\rm B}/\hbar}Be^{-itH_{\rm B}/\hbar};
\label{tartar1}
\EEA
where $S(t)$ and $B(t)$ are free Heisenberg operators of the bath
(i.e.
they evolve under the uncoupled system and bath dynamics),
${\rm h.c.}$ before the end of the curly bracket means the hermitean
conjugate 
of the whole expression contained in this bracket, and where
\BEA
K(t,\theta)=\langle B(t)B(\theta)\rangle\equiv{\rm
tr}\,[B(t)B(s)D_{\rm B}(0)]
\EEA
is the {\it free} correlation function of the bath variables.

Now assume that the decoupling time $\tau$ of the correlation function

$K(t,\theta)$ is the smallest characteristic time of the considered
situation:
\BEA
K(t,\theta)\simeq\langle B(t)\rangle\langle B(\theta)\rangle=0, \quad
{\rm for}\quad |t-s|\gg \tau,
\label{de}
\EEA
where the last equality is realized due to (\ref{abbat}), which in the
present
context reads: $\langle B(t)\rangle=\langle B(\theta)\rangle=0$.
This means that for $t\gg\tau$ the relevant integration domain of the 
integrals over $\theta$ in (\ref{bresult}) is $\theta\ll t$, and the
upper
limit of these integrals can be substituted by infinity. Then the
whole
operator acting on $D_{\rm S}(t)$ can be viewed as $t$-independent.
Thus, the
solution of (\ref{bresult}) is represented as:
\BEA
D_{\rm S}(t)=\exp\left[\,t{\cal L}_{eff}\,\right]\,D_{\rm S}(0),
\EEA
with an effective Liouville operator ${\cal L}_{eff}$ obtained from
(\ref{bresult}).
This is just the desired form (\ref{baba}). Thus, provided that the
stationary distribution $D_{\rm S}^{(st)}$ exists, the properties 
(\ref{baba}, \ref{handi}) are satisfied, and the thermodynamical arrow
of time has been established as folows from Eq.~(\ref{saladin}).

The decoupling property (\ref{de}) is seen to be connected with
dynamics of the free bath, see (\ref{tartar1}), and hence needs a
concrete physical mechanism for its validity.  The most standard
mechanism for this is to take a very large bath, consisting of many
nearly independent pieces.  Another possible mechanism is the
intrinsic chaoticity of the bath, which leads to decoupling of
correlators \cite{mackay,zaslav,j}.  This property will be discussed
in the next section.

\renewcommand{\thesection}{\arabic{section}}
\section{ Geodesics mixing}
\setcounter{equation}{0}\setcounter{figure}{0} 
\renewcommand{\thesection}{\arabic{section}.}
\label{geo}

The geodesics of a space (locally if the space is non-compact) with
constant negative curvature $k$ in all two-dimensional directions
are known to possess properties of Anosov systems.

The Jacobi equation which describes the deviation ${\bf n}$ of close
geodesics
\begin{equation}
\frac{d^2 {\bf n}}{d\lambda^2}+k{\bf n}=0,
\end{equation}
for $k=-1$ has the solution
\begin{equation}
{\bf n}={\bf n(0)} \cosh \lambda + \dot{{\bf n}}(0)\sinh \lambda.
\end{equation}  
It was proved \cite{Po} (see also \cite{Col}) that for a dim=3 compact
manifold $M$ with constant negative curvature
the time correlation function of the geodesic flow $\{f^\lambda\}$
on the unit tangent bundle $SM$ of $M$
\BEA
b_{A_1,A_2}({\lambda})&&=\int_{SM}A_1(f^{\lambda}x) A_2(x) d\mu
\nonumber\\
&&-\int_{SM}A_1(x)d\mu\int_{SM}A_2(x) d\mu
\EEA
decays exponentially for all functions $A_1,A_2\in L^2(SM)$
\begin{equation}
\left|b_{A_1,A_2}(\lambda)\right|
\leq c\cdot \left|b_{A_1,A_2}(0)\right|\cdot e^{-h\lambda} \ ,
\label{expmix}
\end{equation}
where $c>0$, $\mu$ is the Liouville measure and $\mu(SM)=1$,
$h$ is the KS (Kolmogorov-Sinai) entropy of the geodesic flow
$\{f^{\lambda}\}$.
To reveal the properties of the free motion of photons in
pseudo-Riemannian (3+1)-space the projection of its geodesics
into Riemannian 3-space has to be performed, i.e. by corresponding
a geodesic $c(\lambda)=x(\lambda)$ to the geodesic in the former
space:
$\gamma(\lambda)=(x(\lambda),t(\lambda))$.
Then the transformation of the affine parameter is as follows
\cite{LMP}:
$$
\lambda(t)=\int_{t_0}^t \frac{ds}{a(s)} \ .
$$
The KS-entropy in the exponential index can be easily estimated
for the matter-dominated post-scattering Universe \cite{g1},
so that
\begin{equation}
e^{h\lambda}=
(1+z)^2\left[\frac{1
+\sqrt{1-\Omega}}{\sqrt{1+z\Omega}+\sqrt{1-\Omega}}\right]^4 \ .
\label{factor}
\end{equation}
i.e. depends on the density parameter $\Omega$ and the redshift of the
last scattering epoch $z$.
The initial condition (\ref{mis3}) should be
the one to ensure the thermodynamical arrow.
The decay of correlators for geodesic flow for a $k=-1$ FRW Universe
provides the procedure of coarse-graining and ensures the
Markovian (no-memory) behavior of the CMB parameters, so that
\begin{equation}
\label{c1}
t\gg \tau = 1/h,
\end{equation}
where KS-entropy defines the characteristic time scale $\tau$ and
depends only on the diameter of the Universe which is the only scale
in the maximally symmetric space \cite{g1}.  The time scale $\tau$ is
the so-called Markov time or a random variable independent of future
as defined in the theory of Markov processes \cite{Dyn}, and in the
CMB problem describes the decay of initial perturbations, i.e. damping
of the initial anisotropy amplitude and the flattening of the angular
correlation function \cite{g1}, \cite{g2}.  For certain dynamical
systems that time scale defines also the relaxation time for tending
to a microcanonical equilibrium.  The evaluation of $\tau$ and hence
of the negative curvature of the Universe has been performed in
\cite{g3} using the COBE data.  The negative constant curvature leads
to a decay of time correlators of geodesics, thus defining the
thermodynamic arrow for CMB in a FRW $k=-1$ Universe.

\renewcommand{\thesection}{\arabic{section}}
\section{Gibbs distribution, the second law and the arrow of time.}
\setcounter{equation}{0}\setcounter{figure}{0} 
\renewcommand{\thesection}{\arabic{section}.}

The purpose of the present section is to discuss to what extent
irreversibility, and the thermodynamical arrow of time are {\it
necessary} to establish the second law, and the Gibbs distribution,
which are known to be the basis of equilibrium statistical mechanics.
The development of statistical thermodynamics during more than one
century safely confirms the sufficiency of the thermodynamical arrow
of time to derive the Gibbs distribution and the second law
\cite{landau,jaynes}.  To show that they are not necessary, we shall
consider an alternative derivation of the Gibbs distribution proposed
by Lenard \cite{Lenard}. Similar ideas were expressed in \cite{W}.
For a recent extension of these results see \cite{ATheo}.

A closed statistical system is considered. Its dynamics is described
by a Hamiltonian $H$. At the moment $t=0$, where the state of the
system is $D(0)$, an external time-dependent field is switched on, and
the Hamiltonian becomes ${\cal H}(t)$. The field is switched off at
the moment $t$, and the Hamiltonian will again be $H$ (cyclical
variation).  The following postulate is imposed: It is impossible to
extract work from the system in the state $D$ by switching any
external field in such a way.  This is the statement of the second law
in the Thomson formulation \cite{ATheo}. We shall see that this
condition alone plus some companion ones are enough to derive the
Gibbs
distribution for the density matrix $D$ of this system. Notice that
the presented condition does not impose irreversibility. Indeed, the
evolution of the system remains purely unitary (thus reversible).  As
the result of the time-dependent field an external source has done the
work \cite{balian} 
\BEA
\label{work}
W&&=\int^{t}_{0}d\theta~{\rm tr}[D(\theta)\frac{d{\cal
H}(\theta)}{d\theta}]
\\
&&={\rm tr}[H(D(t)-D(0))],
\label{dudki}
\EEA
where, when going from (\ref{work}) to (\ref{dudki}),
we used integration by parts, and the equation of motion
\begin{equation}
\label{motion}
i\hbar\dot{D}=[{\cal H}(t),\rho (t)].
\end{equation}
Let us now introduce a unitary operator
$V(t)$,
\begin{equation}
\label{7}
D(t)=e^{-itH/\hbar}VD(0)V^{\dagger}e^{itH/\hbar},
\end{equation}
and rewrite Eq.~(\ref{work}) as 
\begin{equation}
\label{8}
W={\rm tr}[D(0)V^{\dagger}HV]-{\rm tr}[D(0)H].
\end{equation}
The quantity $W$ is required to be positive. Since $W=0$ for
$V=1$, we have to demand that ${\rm tr}[D(0)V^{\dagger}HV]$
is minimal for $V(1)=1$. For $V$ close to $1$ one introduces an
expansion
\BEA
V=1+M+{\cal O}(M^2)
\EEA
where $M^{\dagger}=-M$, and $M$ is small. One obtains
\BEA
W={\rm tr}([D(0),H]M)+{\cal O}(M^2).
\EEA
Since the sign of ${\rm tr}([D(0),H]M)$ can be arbitrary,
$D(0)$ and $H$ should commute for $W$ to be minimal at $V=1$. 
To obtain a more precise relation between the eigenvalues, we use 
a particular form 
\begin{equation}
\label{33}
V=\left [\begin{array}{rr}
\cos \theta & \sin \theta \\
-\sin \theta & \cos \theta
\end{array}\right ],
\end{equation}
when acting on the two-dimensional subspace formed by 
common eigenvectors $|i\rangle$ and $|k\rangle$ of $H$ and $D(0)$, 
and
$V=1$ in the orthogonal complement of this subspace. Now it easy to
obtain
\BEA
W=-(h_i-h_k)(r_i-r_k)\sin ^2 \theta,
\EEA
where $h_i$ and $h_k$ (or $r_i$ and $r_k$) are the corresponding
eigenvalues
of $H$ (or $D$). It is seen for
any $|i\rangle$, $|k\rangle$ that 
\begin{equation}
\label{35}
{\rm if}\, \, h_i\ge h_k,\,\, {\rm then}\,\, r_i\le r_k.
\end{equation}
As follows from this, there is some positive non-increasing function
$f$,
such that 
\BEA
\label{79}
D =f(H).  \EEA 
The exponential form of $f$ can be established from the
standard reasons of extensivity. It is assumed that the form of the
function $f$ is (at least to some extent) universal, and does not
depend on the Hamiltonian itself. Then for two interacting subsystems
we have

\begin{eqnarray}
\label{100}
&&{\rm lim}_{g\to 0}~f(H_1+H_2+gH_{int})
\nonumber \\
&&=f(H_1+H_2)=f(H_1)~f(H_2),
\end{eqnarray}
where $gH_{int}$ is the Hamiltonian of the interaction.
Under reasonable conditions $f$ can be proven to be exponential:
\BEA
\rho (H)=\frac{1}{Z}\exp (-\beta H),
\EEA
with $\beta =1/T\ge 0$. 

\renewcommand{\thesection}{\arabic{section}}
\section{ Conclusion}
\setcounter{equation}{0}\setcounter{figure}{0} 
\renewcommand{\thesection}{\arabic{section}.}

The main purpose of the present paper is to suggest a new viewpoint on
possible
connections between the thermodynamical and cosmological arrows of
time. 
One aspect of such connections is well-known and is based on the
important role
of special initial conditions for both arrows
\cite{Penrose1,Penrose2}.
However, the special initial conditions are by no means sufficient for

generating the thermodynamical arrow of time: they are able to
generate a
pre-arrow of time only. Thus initial conditions cannot be the only
part of 
the story. We show that the negative curvature of the 
Friedmann-Robertson-Walker Universe and the effect of
geodesic mixing 
can provide the condition
necessary for the emergence of the thermodynamic 
arrow of time.
Moreover, this mechanism can explain why CMB contains the
major fraction of the entropy of the Universe.

If this is indeed an origin of the thermodynamic arrow,
then the thermodynamics in flat and positively curved universes need
not
be strongly time asymmetric, and the latter is observed
since we happen to live in a Universe with negative curvature.
In particular, as we show in section IV, this does not mean
that the second law of thermodynamics and the Gibbs distribution will
have 
less chances to survive for non-negatively curved spaces, since these
concepts do not necessarily depend on the thermodynamical arrow of
time.

In other words, the symmetry of the Newtonian mechanics,
electrodynamics, quantum mechanics and equilibrium statistical 
thermodynamics might purely survive in some universes.
In this context the essence of thermodynamical arrow must be
understood as
not the mere
increase of entropy of an almost closed system, but the fact that this
arrow has the universal direction in the entire Universe (see
\cite{GMH}).
In the light of our suggested explanation
of the emergence of this arrow, it may follow that the negative
curvature is
the very mechanism unifying all local thermodynamical arrows.
According to 
this logic in the flat or positively curved universes, i.e. at the
absence 
of a global unification mechanism, there can be local thermodynamical
arrows
with various directions.

Another intriguing problem arising here, is whether life can occur in
such globally time-symmetric universes, or the time asymmetry/negative
curvature is a necessary ingredient for developing of life | the {\it
curvature anthropic principle}. The CMB has to carry the signature of
this principle.

\section*{Acknowledgments}
We are grateful to the anonymous referee for an unusual care in the
refereeing
and editing of the paper.

\end{document}